%% file: paper.tex
\documentclass[letterpaper]{article} 
\usepackage{aaai25}  
\usepackage{times}  
\usepackage{helvet}  
\usepackage{courier}  
\usepackage[hyphens]{url}  
\usepackage{graphicx} 
\usepackage{natbib}  
\usepackage{caption} 
\usepackage{algorithm}
\usepackage{algorithmic}
\usepackage{booktabs} 
\usepackage{colortbl}
\usepackage{xcolor}
\usepackage{graphicx} 
\usepackage{newfloat}
\usepackage{listings}

\usepackage{cuted}

\DeclareCaptionStyle{ruled}{labelfont=normalfont,labelsep=colon,strut=off} 
\floatstyle{ruled}
\newfloat{listing}{tb}{lst}{}
\floatname{listing}{Listing}
\usepackage{amsmath}
\usepackage{amssymb}

\title{Revisiting CAD Model Generation by Learning Raster Sketch}
\author{
    Pu Li\textsuperscript{\rm 1,\rm 2}\equalcontrib,
    Wenhao Zhang\textsuperscript{\rm 1}\equalcontrib,
    Jianwei Guo\textsuperscript{\rm 3,\rm 1},
    Jinglu Chen\textsuperscript{\rm 1},
    Dong-Ming Yan\textsuperscript{\rm 1,\rm 2}\thanks{Corresponding author.}
}

\affiliations{
    \textsuperscript{\rm 1}MAIS, Institute of Automation, Chinese Academy of Sciences\\
    \textsuperscript{\rm 2}School of Artificial Intelligence, University of Chinese Academy of Sciences \\
     \textsuperscript{\rm 3}School of Artificial Intelligence, Beijing Normal University

    \{lipu2021, wenhao.zhang, chenjinglu\}@ia.ac.cn, jianwei.guo@bnu.edu.cn, yandongming@gmail.com
}

\usepackage{bibentry}

\begin{document}

\maketitle
\begin{abstract}
The integration of deep generative networks into generating Computer-Aided Design (CAD) models has garnered increasing attention over recent years. Traditional methods often rely on discrete sequences of parametric line/curve segments to represent sketches. Differently, we introduce RECAD, a novel framework that generates \textbf{R}aster sketches and 3D \textbf{E}xtrusions for \textbf{CAD} models. Representing sketches as raster images offers several advantages over discrete sequences: 1) it breaks the limitations on the types and numbers of lines/curves, providing enhanced geometric representation capabilities; 2) it enables interpolation within a continuous latent space; and 3) it allows for more intuitive user control over the output. Technically, RECAD employs two diffusion networks: the first network generates extrusion boxes conditioned on the number and types of extrusions, while the second network produces sketch images conditioned on these extrusion boxes. By combining these two networks, RECAD effectively generates sketch-and-extrude CAD models, offering a more robust and intuitive approach to CAD model generation. Experimental results indicate that RECAD achieves strong performance in unconditional generation, while also demonstrating effectiveness in conditional generation and output editing.
\end{abstract}

\section{Introduction}

\begin{figure}[t]
\centering
\includegraphics[width=1\linewidth]{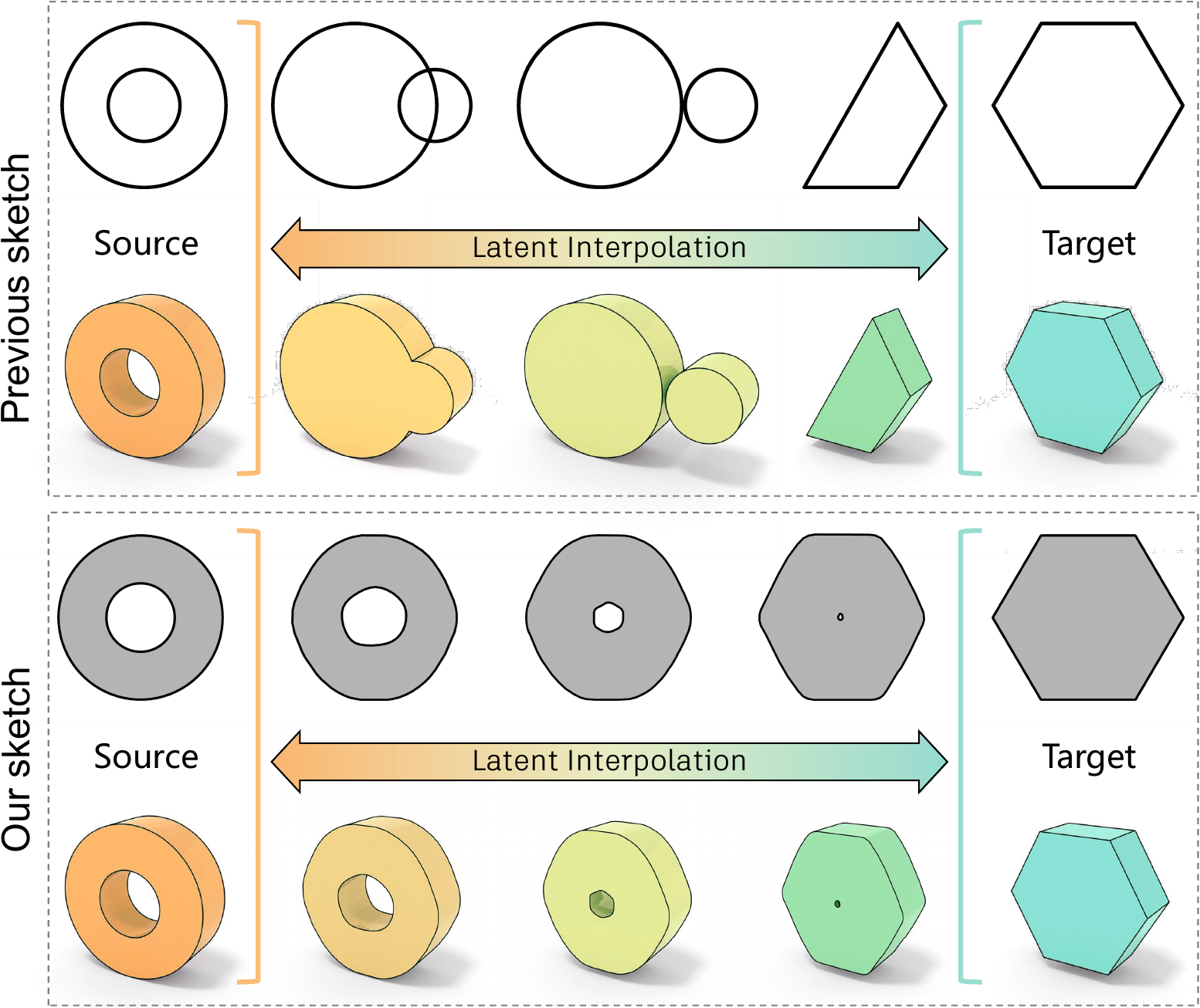}
\caption{Top: Previous approaches \cite{wu2021deepcad,xu2022skexgen,xu2023hierarchical,wang2024vq} represent sketches using vector sequences, which often fail to produce smooth transitions in shape and topology during latent space interpolation. Bottom: RECAD learns from rasterized sketches, ensuring more plausible interpolated shapes, which results in more natural and continuous transformations between different sketches. }\label{fig:teaser}
\end{figure}

The digital genesis of modern artifacts, from everyday consumer products to complex industrial machinery, is now deeply intertwined with Computer-Aided Design (CAD) systems. Central to many CAD workflows is sketch-based modeling, where 2D sketches imbued with geometric constraints and design intent are transformed into intricate 3D models through a series of feature-based modeling operations, ultimately giving rise to complex assemblies.

Among these feature-based modeling operations, extrusion is the most prevalent, allowing designers to generate 3D shapes by extending 2D sketches along a defined path.  However, the traditional parametric approach to sketch extrusion, while powerful, presents significant challenges. Creating these parametric models, where 2D sketches are extruded into 3D forms based on numerous parameters and equations, is tedious and expert-intensive. Furthermore, the rigid nature of these parametric dependencies makes major design modifications difficult, often causing models to fail. This lack of flexibility hinders iterative design and highlights the need for more intelligent CAD systems. 

\begin{figure}[t]
\centering
\includegraphics[width=1\linewidth]{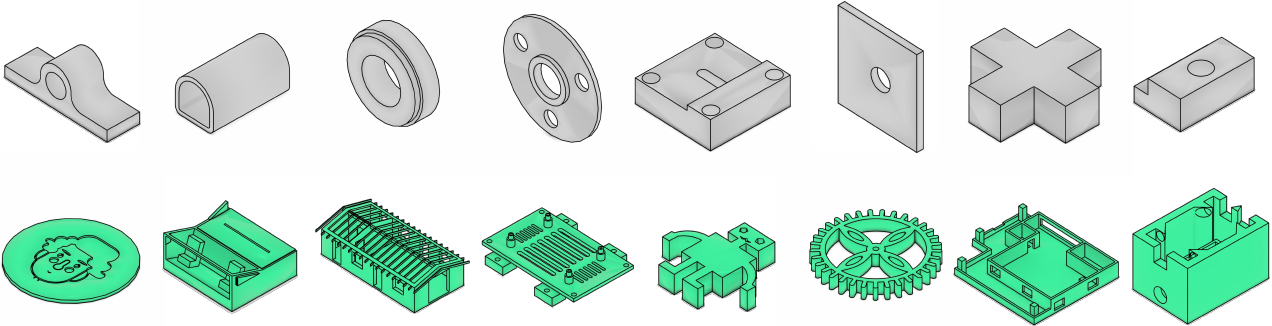}
\caption{Sample data from DeepCAD dataset. Previous methods, constrained by sequence length limitations, were restricted to learning a subset of simpler shapes within the dataset (gray). Our raster-based sketch representation enables the learning of more complex shapes (green).}
\label{fig:dataset}
\end{figure}

With the advance of deep neural networks, learning-based methods have emerged as a promising avenue for generating diverse CAD models. A prevalent approach involves training generative models by leveraging construction sequences as supervision \cite{wu2021deepcad, xu2022skexgen, xu2023hierarchical, wang2024vq}. These methods typically represent sketch-and-extrude sequences as parameterized discrete sequences, enabling the application of powerful sequence learning techniques to the task of CAD model generation.

Despite the demonstrated success of sequence-based learning methods for CAD model generation, several limitations remain: 1) The reliance on sequences introduces a bottleneck, limiting the complexity of representable sketches and the number of commands that can be effectively utilized. This constraint restricts the generation of intricate models. 2) Current methods often generate semantically invalid command sequences for CAD modeling, resulting in non-viable models and hindering the reliability and usability of the output. 3) Representing sketches as discrete parameter sequences makes generating smooth and continuous shape transformations difficult, leading to unnatural results for latent space interpolation, as illustrated in Fig.\ref{fig:teaser}. This poses a challenge for generating plausible sketches, for which plausible interpolation is essential \cite{radford2015unsupervised, goodfellow2014generative, higgins2017beta}.

In this paper, we revisit the CAD generation task and propose a novel representation that follows the sketch-and-extrude paradigm without using command sequences. Instead of representing sketches as curve sequences, we leverage raster images. Extrusion parameters, such as origin, direction, and Boolean operation type, are implicitly determined by the dimensions and type of the extrusion box. This new representation addresses the limitations of sequence-based methods in several key ways. 
Using raster images to represent sketches allows for depicting highly complex shapes, overcoming the bottleneck imposed by discrete parameter sequences (e.g., fixed maximum length of curves per loop), as shown in Fig.\ref{fig:dataset}.
Additionally, the intuitive nature of image-based sketches reduces the likelihood of generating semantically or geometrically invalid commands, leading to more reliable and usable outputs. Moreover, representing sketches as images facilitates the generation of smooth and continuous shape transformations, unlike the discrete nature of previous methods. Finally, by employing extrusion boxes, our method provides a concise and intuitive way to control the topological relationships between extruded shapes, further enhancing the modeling process.

Building upon this novel sketch-extrusion representation, we present RECAD, a novel generative framework that generates sketch-extrusion pairs using Denoising Diffusion Probabilistic Models (DDPM) \cite{ho2020denoising}. Concretely, RECAD comprises a sketch image VAE and two diffusion networks. A Variational Autoencoder (VAE) \cite{kingma2013auto} is utilized to learn the latent features of raster sketches. Two diffusion networks are employed in a sequential generation process. The first diffusion network generates the coordinates defining the extrusion boxes. Conditioned on these boxes, the second network generates latent features representing the raster sketches. By extruding sketches into 3D solids and integrating the topology defined by the extrusion boxes, we achieve the final CAD model.

Leveraging the adaptable architecture of RECAD, our work explores advancements in controllable generation. Beyond the typical autocompletion capabilities, we introduce a novel feature: extrusion box conditional generation. This allows for automatically generating sketches based on a user-defined structure of extrusion boxes, resulting in a detailed 3D model. Additionally, RECAD allows users to easily edit generated sketches using simple image editing tools, such as monochrome brushes, to modify local geometry in a natural and intuitive manner. In summary, our contributions include:
\begin{itemize}
    \item A novel sketch-extrusion representation leverages raster sketch and extrusion box pairs for CAD model construction.
    \item A latent diffusion generative framework that enables random generation of complicated CAD models, auto-completion from partial input, extrusion box conditional generation, and flexible user editing.
    \item Extensive experiments demonstrate the superiority of RECAD in CAD generation through comprehensive comparisons with existing state-of-the-art approaches.
\end{itemize}

\section{Related Work}

\subsubsection{Constructive solid geometry.} 
Constructive Solid Geometry (CSG) represents complex 3D shapes through Boolean operations on simple geometric primitives.
This compact representation has been employed for CAD reconstruction tasks with program synthesis \cite{du2018inversecsg, nandi2017programming, nandi2018functional}. 
For instance, CSGNet \cite{sharma2018csgnet} uses reinforcement learning to minimize reconstruction errors. UCSGNet \cite{kania2020ucsg} takes this a step further by eliminating the need for ground truth CSG trees during inference. Other works also explored unsupervised learning with specialized parametric primitives \cite{chen2020bsp, ren2021csg, yu2022capri, yu2023dualcsg}. 
While deep learning methods hold promise for advancing CSG-based 3D model generation, they are still in the exploration phase, with limited datasets posing a significant challenge.

\subsubsection{B-rep CAD generation.}
Recent research endeavors have sought to directly generate Boundary Representation (B-rep) CAD models. 
Direct B-rep generation involves constructing the necessary geometric and topological elements, trimming surface patches, and joining them into a unified solid model. Leveraging learning-based methods, previous works have made advances in generating parametric curves \cite{wang2020pie} and surfaces \cite{sharma2020parsenet,li2023surface}. To replace fixed topological templates \cite{smirnov2019learning}, creating generalized topology for sketches \cite{willis2021engineering} and solid models \cite{wang2022neural, guo2022complexgen, jayaraman2022solidgen, xu2024brepgen} is then explored. 

\subsubsection{Sketch-and-extrude CAD generation.} 
Sequential CAD modeling operations combined with learning-based methods have made significant progress on several tasks. Some works progress in generating 2D engineering sketches \cite{willis2021engineering, para2021sketchgen, ganin2021computer, seff2021vitruvion} in editable parametric CAD files obtained from the generated sequences, and converting between command sequences and hand drawings \cite{li2020sketch2cad, li2022free2cad, hahnlein2022cad2sketch}. DeepCAD \cite{wu2021deepcad} produces a large-scale dataset of CAD models in the form of sketch-and-extrude sequences and first learns to generate editable CAD models. Later works further improve the performance of shape generation by employing auto-regressive generative models \cite{xu2022skexgen}, designing a hierarchical tree of neural codes mastering both high level concepts and local level geometry \cite{xu2023hierarchical}, and integrating discrete diffusion models \cite{wang2024vq}. In the field of 3D model reconstruction, approaches learning 2D sketches and predicting the corresponding extrusions to reconstruct the CAD models in supervised \cite{uy2022point2cyl}, and unsupervised \cite{li2023secad,ren2022extrudenet, li2024sfmcad} manners are proposed. 

This work explores CAD model generation, focusing specifically on the sketch-and-extrude modeling paradigm. Departing from prior art that relies on discrete parametric representations of sketches, we propose leveraging raster images as a more flexible and expressive alternative. Furthermore, we encapsulate all extrusion information within an extrusion box.

\subsubsection{Generative diffusion models.}
DDPM \cite{ho2020denoising} are good at generative tasks, especially in image generation and restoration \cite{ho2020denoising, saharia2022image, lugmayr2022repaint}. Latent diffusion models (LDM) operate in a low-dimensional latent space rather than original data space, resulting in more efficient sampling and promising generative performance on numerous tasks such as high-resolution image synthesis \cite{vahdat2021score, rombach2022high}. Diffusion models also show the ability to generate geometry with \cite{shabani2023housediffusion, liu2024cage} or without topology as input conditioning \cite{alliegro2023polydiff, xu2024brepgen}. Borrowing architecture from VQ-Diffusion \cite{gu2022vector} which combines VQ-VAE \cite{van2017neural} and conditional DDPM, VQ-CAD \cite{wang2024vq} integrates discrete diffusion models for CAD model generation. Our work includes two diffusion networks for generating 3D extrusion boxes and raster sketches representing detailed 2D geometry. 

\section{Representation and Overview}

\begin{figure*}[!t]
\centering
  \includegraphics[width=1\linewidth]{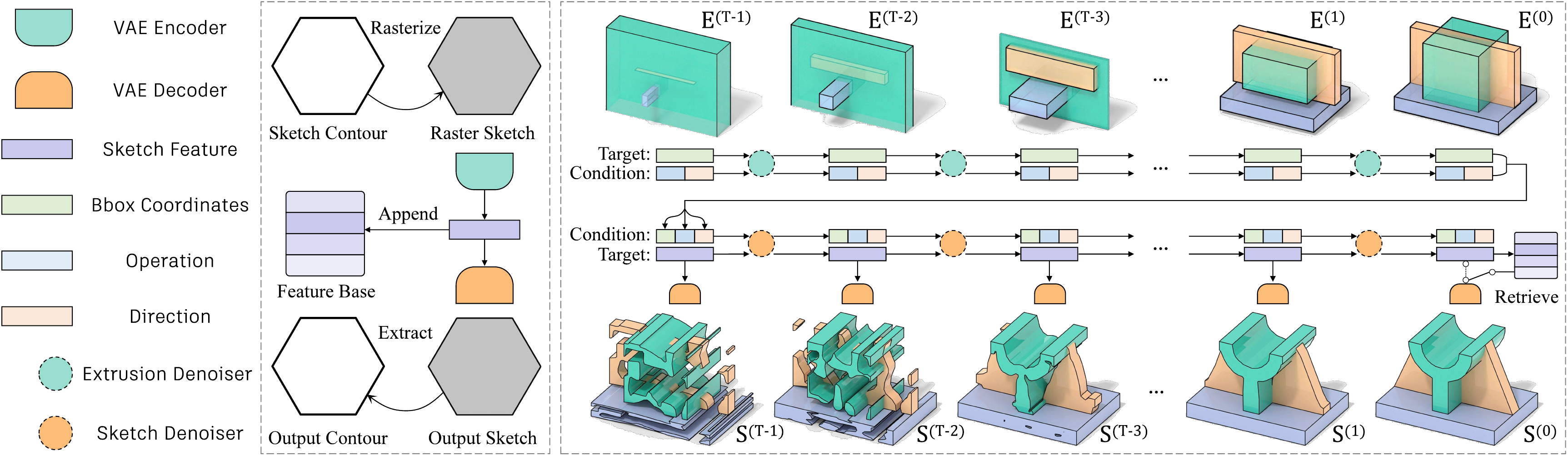}   
  \caption{Framework of RECAD. A sketch image VAE, shown on the left, learns a latent feature base of rasterized input sketches. This feature base is then utilized for efficient sketch retrieval. The right panel details the two-stage diffusion process. First, extrusion boxes are progressively generated conditioned on user-specified operation and direction. Subsequently, sketch features are generated conditioned on these bounding boxes and then decoded into raster sketches using the VAE decoder. E$^{(t)}$ and S$^{(t)}$ represent visualizations of the extrusion boxes and the sketch-extrusion result at step $t$, respectively.}
  \label{fig:network}
\end{figure*}

In this section, we first introduce our novel representation using raster sketches paired with extrusion boxes. We then provide an overview of our proposed approach for generating CAD models based on this representation.

\subsection{Novel Sketch-Extrusion Representation}

Sequence-based representations, while commonly used in CAD model generation, limit design complexity, semantic validity, and smooth shape transformations. To address these challenges, we propose a novel sketch-extrusion paradigm that leverages raster images for sketch representation, allowing for greater complexity and smoother transformations. Additionally, we introduce extrusion boxes, establishing a direct mapping between the 2D sketch and the 3D extrusion. This representation simplifies the extrusion parameters themselves into a few key controls within the extrusion box, allowing for intuitive manipulation of the resultant topology. Below, we detail our raster sketch representation and the extrusion box concept.

\subsubsection{Raster sketch.}
A raster sketch is a binary image depicting the shape of the sketch. Pixel values of 1 correspond to areas suitable for extrusion, while 0 represents empty space. This image-based representation allows for the portrayal of complex geometries without concern for the count of internal loops and curves. Moreover, compared to the sequence-based representation, this pictorial format facilitates more straightforward and valid interpolation within the latent space. 

\subsubsection{Extrusion box.}
The DeepCAD dataset defines extrusion using nine continuous and two discrete parameters. We note that the extrusion direction for most models aligns with the principal axes. Furthermore, the symmetry of extrusion allows us to consolidate bidirectional extrusion into a unidirectional representation. Motivated by these observations, we propose a simplified and more straightforward representation called the  ``extrusion box". An extrusion box is fully defined by eight parameters:
\begin{equation}
    E = (x_{min}, y_{min}, z_{min}, x_{max}, y_{max}, z_{max}, \omega, \mathcal{B}).
\end{equation}
The first three parameters represent the bottom-left corner of the extrusion box, while the subsequent three indicate the top-right corner. The term $\omega$ specifies the extrusion direction along one of the principal axes (i.e., x, y, or z). $\mathcal{B}$ denotes one of the three Boolean operations (i.e., union, subtraction intersection, or intersection). 
This representation directly couples the 2D sketch to the 3D extrusion: The dimensions of the sketch are determined by the ranges of the two axes perpendicular to the extrusion direction, while the range along the extrusion direction dictates the extrusion height.

Each pair of raster sketch and extrusion box can naturally form a solid through the sketch-extrude process. Multiple sketch-extrusion pairs can link the generated solids topologically to express complex CAD models.

\subsection{Overview}
We aim to generate sketch-extrusion pairs in the format of our novel representation and convert them into parametric CAD models. In RECAD, a sequential two-stage denoising is utilized for shape generation, as illustrated in Fig.\ref{fig:network}. In the first stage, a set of extrusion topology guidance $E_g$ comprising one-hot encoded direction and operation is randomly sampled from its ground-truth distribution. We then use a Transformer-based denoiser to denoise a random Gaussian noise conditioned on $E_g$, and obtain the extrusion coordinates $E_c$. A set of complete extrusion boxes $E$ are formed by combining $E_g$ and $E_c$. In the second stage, a UNet-based denoiser is employed to generate latent sketches $S_z$ conditioned on the complete extrusion boxes $E$. The latent sketches $S_z$ are then decoded to sketch images $S$ through a sketch image VAE.
We finally combine the generated extrusion boxes and sketch images to construct CAD models.

\section{Method}

Here, we present RECAD, a novel generative framework comprising a sketch image VAE and two diffusion networks.
\subsection{Sketch Image VAE}

We employ a sketch image VAE to compress sketch images into latent embeddings as in Stable Diffusion \cite{rombach2022high}. The backbone of the sketch image VAE is a 2D convolution UNet \cite{ronneberger2015u}. The grayscale images to be used as VAE input are all resized to a size of $32\times32$. Concretely, given a binary sketch image $S \in \mathbb{R}^{32\times32\times1}$, the VAE encoder $\mathcal{E}$ encodes $S$ into a latent representation by a downsampling factor of 8 and a feature depth of 3, yielding $S_z \in \mathbb{R}^{4\times4\times3}$. The VAE decoder $\mathcal{D}$ then reconstructs the sketch image $\tilde{S}=\mathcal{D}(S_z)=\mathcal{D}(\mathcal{E}(S))$. The training process of the sketch image VAE involves optimizing two components: the mean squared error (MSE) for reconstruction loss between $S$ and $\tilde{S}$, and the Kullback-Leibler (KL) divergence as a regularization term. 

\subsection{Two-Stage Diffusion}

Diffusion models are generative models that learn to gradually denoise a sample from Gaussian noise towards a data distribution, consisting of a forward diffusing process that adds noise and a reverse denoising process that adds structure. Our two-stage diffusion contains a Transformer-based extrusion denoiser and a UNet-based sketch denoiser. We train the denoisers with topologically organized extrusion boxes and latent sketches with inherent geometry details, following the training strategy in DDPM \cite{ho2020denoising}. 

\subsubsection{Forward diffusing process.}
During training, the forward process takes a data sample $x_0$, which could be extrusion coordinates or latent sketch features in our case, and gradually adds Gaussian noise to the data in T time steps. Each forward step is computed by:
\begin{equation}
    q(x_t | x_{t-1}) = \mathcal{N}\left(x_t; \sqrt{1 - \beta_t} x_{t-1}, \beta_t \mathbf{I}\right), 
\end{equation}
where $\beta_t$ is a noise variance schedule. At a specific time step $t$, the noisy sample features $x_t$ are sampled as
\begin{equation}
    x_t = \sqrt{\bar{\alpha}_t} x_0 + \sqrt{1 - \bar{\alpha}_t} \epsilon_t
\label{eq: x_t}
\end{equation}
using the notation $\bar{\alpha}_t = \prod_{i=1}^{t} \alpha_i$, ${\alpha}_t = 1 - \beta_t$ and Gaussian noise $\epsilon_t \sim \mathcal{N}(0, \mathbf{I})$. After T time steps, the data sample will be completely corrupted into a random noise. The latent sketches will lose their inherent shape information, and the coordinate vectors of extrusions will be scrambled, resulting in the forgetting of relative topology.

\subsubsection{Two-stage denoising process.}
We then train DDPM to reverse the diffusing process. The reverse process denoises a pure Gaussian noise step by step until recovering $x_0$ and learns neural networks to directly predict the noise-removed data distribution of extrusion boxes and latent sketches. 

We apply a sequential two-stage denoising, which first generates extrusion boxes conditioned on ground-truth extrusion guidance and then generates latent sketches conditioned on these extrusion boxes.  

At the beginning of the first stage, we randomly sample a set of extrusion guidance $E_g\in \mathbb{R}^{5}$ from its ground-truth distribution $E_g^*$, which is derived from a histogram of extrusion directions and Boolean operations in our training data:
\begin{equation}
    E_g = (\Phi(\omega, 3) ,\Phi(\mathcal{B}, 2))\sim P(E_g^*),
\end{equation}
where $\Phi$ denotes one-hot encoding. Boolean operation $\mathcal{B}$ is encoded as a 2D vector, encompassing only union and difference.
The vector $E_g$ acts as the conditioning input for the denoiser with a Transformer backbone \cite{vaswani2017attention}, which processes the noisy extrusion box coordinates $E_c \in \mathbb{R}^{6}$. $E_g$ and $E_c$ are embedded with the same feature dimension:
\begin{equation}
    \mathbf{E_g} = \text{MLP}(W_gE_g), \mathbf{E_c} = \text{MLP}(W_cE_c),
\end{equation}
where $W_g \in \mathbb{R}^{n\times5}$ and $W_c \in \mathbb{R}^{n\times6}$ are embedding matrices of $n$-dimension. MLP is fully connected layers with SiLU activation. We employ direct condition injection in this stage: $\mathbf{\breve{E}_c} \leftarrow \mathbf{E_c} + \mathbf{E_g}$.
The first stage produces a set of denoised extrusion coordinates, $E_c^{de}\in\mathbb{R}^{6}$. These coordinates are concatenated with the corresponding extrusion guidance, $E_g$, forming the complete extrusion boxes, $E \in \mathbb{R}^{11}$:
\begin{equation}
    E = E_c^{de} \: \Vert \: E_g.
\end{equation}
$E$ serves as the condition for the second denoiser. 

In the second stage, inspired by the exceptional performance of video diffusion models \cite{ho2022video} in generating sequential images, we similarly employ a 3D conditional UNet \cite{cciccek20163d} as the backbone of the denoiser for sketch sequence generation. Each extrusion box $E$ is embedded as $\mathbf{E} = \text{MLP}(WE)$ where $W \in \mathbb{R}^{d\times11}$ likewise as in the first stage. The input to the second denoiser is latent sketch features $S_z \in \mathbb{R}^{4\times4\times3}$.  In this stage, we apply cross-attention to control the denoising UNet with $\mathbf{E}$ as a conditioning feature. A set of sketch-extrusion pairs are obtained at the end of the two-stage denoising.

Note that time embedding is also passed to the denoiser in each stage. Overall, the distribution of generated extrusion-sketch pairs can be defined as the product of conditional distributions $p(E_c | E_g)$ and $p(S_z | E)$.

\begin{figure*}[!t]
\centering
  \includegraphics[width=1\linewidth]{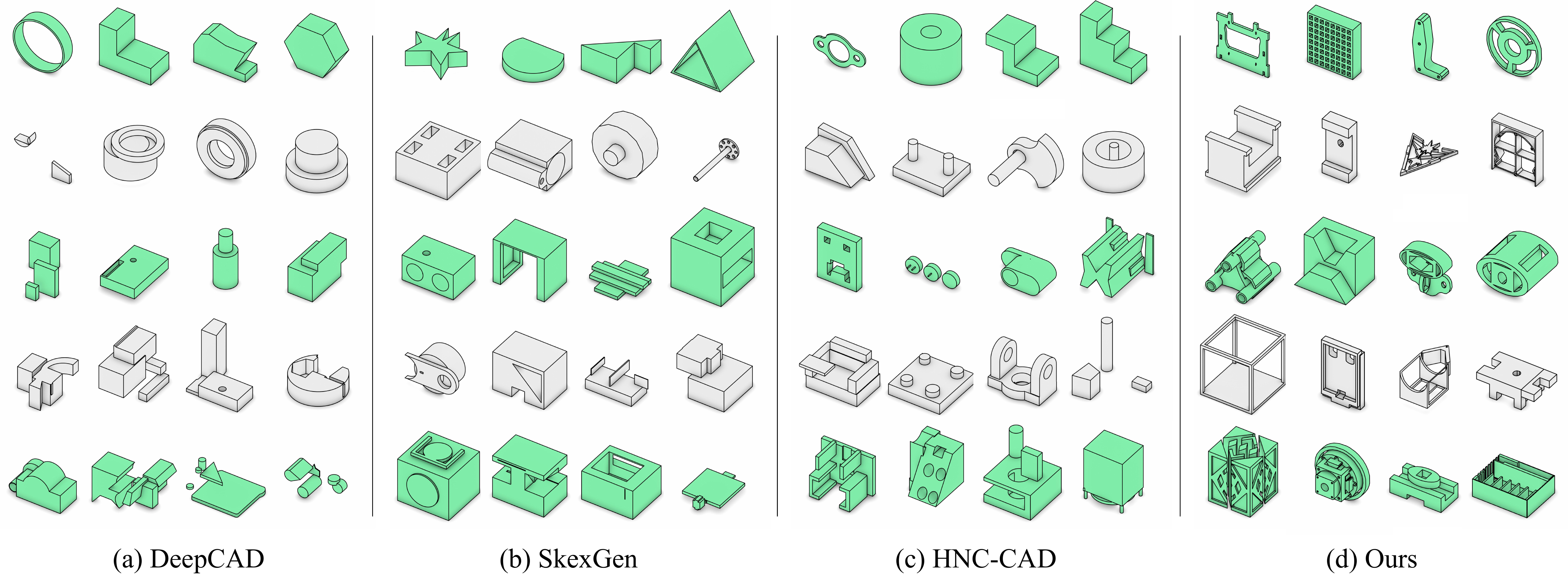}   
  \caption{Visual comparison on the DeepCAD dataset. Each row, from top to bottom, shows shapes generated with 1 to 5 extrusion operations, respectively. Different colors are used for clear visual distinction.}
  \label{fig:comparison}
\end{figure*}

\subsubsection{Loss function.}
Following DDPM \cite{ho2020denoising}, we train our Transformer-based denoiser and 3D UNet-based denoiser respectively with the same simple L2-norm regression loss,
\begin{equation}
    L = \mathbb{E}_{t, x_0, \epsilon_t} \left[ \left\| \epsilon_t - \epsilon_{\theta} \left( \sqrt{\bar{\alpha}_t} x_0 + \sqrt{1 - \bar{\alpha}_t} \epsilon_t, t \right) \right\|^2 \right],
\end{equation}
where $\epsilon_{\theta} \left( \sqrt{\bar{\alpha}_t} x_0 + \sqrt{1 - \bar{\alpha}_t} \epsilon_t, t \right)$ can be simplified to $\epsilon_{\theta} \left( x_t, t \right)$ using Eq. \ref{eq: x_t}. $\epsilon_t$ is the ground-truth Gaussian noise at time step t from the forward diffusing process and is compared against $\epsilon_{\theta} \left( x_t, t \right)$. During sampling, the inferred noise at each time step is utilized to progressively denoise a random Gaussian noise, refining it towards a cleaner signal.

\subsection{Extrusion with Raster Sketch}
The conversion from raster sketches to contours is essential for extruding them into 3D solids. We present two methods for this conversion: 

\subsubsection{Contour extraction.}
We utilize the Teh-Chin chain approximation \cite{teh1989detection} to convert binary raster images into polygonal contours. To streamline the CAD generation process, we simplify these contours using the Douglas-Peucker algorithm \cite{douglas1973algorithms}. The simplified contours are then extruded to 3D prisms.

\subsubsection{Sketch retrieval.}
Our sketch image VAE embeds each training sketch into a feature space. During testing, we retrieve the closest matching sketch from the database based on the Euclidean distance between the generated feature vector of the sketch and the database entries. This retrieved sketch provides predefined contour curves, which are then scaled and extruded to form the 3D solid. Compared to directly extracting contours, this method may limit the diversity of the sketch contours. However, it offers cleaner and more coherent curves, bypassing the process of contour extraction and simplification, thus improving efficiency. We employ sketch retrieval for all experiments except those involving user edits.

\subsection{Implementation Details}

Models are implemented in PyTorch and trained on four NVIDIA RTX A6000 GPUs. We use an AdamW \cite{loshchilov2017decoupled} optimizer with the learning rate 5e-4 for optimization. The sketch Image VAE is trained for 500 epochs at batch size 512 and two denoisers are trained for 2000 epochs at batch size 256. During inference, PNDM \cite{liu2022pseudo} is employed for fast sampling. More details are provided in the supplementary material.

\section{Experiments}

In this section, we demonstrate the effectiveness of our method on two kinds of generation tasks: 1) unconditional generation and 2) controllable generation including autocompletion from partial input, extrusion box conditional generation, and image-based sketch editing.

\subsection{Setup}

\subsubsection{Dataset preparation.} 
We utilize the extensive DeepCAD dataset \cite{wu2021deepcad} which contains CAD models with detailed sequences of sketch-and-extrude modeling operations. 
We follow the procedures of prior works \cite{willis2021engineering, xu2022skexgen, xu2023hierarchical} to detect and eliminate duplicate models within the training set. 

\subsubsection{Evaluation metrics.}

Three metrics evaluate the discrepancy between generated models and the reference test set \cite{wu2021deepcad}.
\begin{itemize}
    \item \textit{Coverage (COV)} measures the diversity of generated shapes by calculating the percentage of ground-truth models that are matched by at least one generated shape using Chamfer Distance (CD). 
    \item \textit{Minimum Matching Distance (MMD)} evaluates the fidelity of generated shapes by averaging the CD between each ground-truth model and its closest generated match. 
    \item \textit{Jensen-Shannon Divergence (JSD)} reports the similarity between the distributions of the reference test set and generated set. 
\end{itemize}

To measure the robustness and uniqueness of generated results, the following metrics are computed \cite{xu2024brepgen}:
\begin{itemize}
    \item \textit{Valid} is the percentage of raw model outputs successfully converted into valid Boundary Representations.
    \item \textit{Novel} is the percentage of generated data that do not exist in the training set. 
    \item \textit{Unique} is the percentage of generated data found only once in the generated set.
\end{itemize}

\subsection{Unconditional CAD Generation}

In this section, we evaluate our models on unconditional CAD generation task. Sketch-extrusion pairs are generated by our diffusion models from random Gaussian noise and then converted to CAD models. We compare our generated samples with state-of-the-art methods, including DeepCAD \cite{wu2021deepcad}, SkexGen \cite{xu2022skexgen}, and HNC-CAD \cite{xu2023hierarchical}. We generate 10,000 CAD models using each method and evaluate them against 2,500 ground-truth models randomly sampled from the reference test set in each run. 
We repeated this evaluation process three times and averaged the metric values across these runs to obtain the final results.

\begin{table}\centering
\resizebox{\columnwidth}{!}{%
\begin{tabular}{lcccccc}
\toprule
Method & COV & MMD & JSD & Valid & Novel & Unique \\ 
       & $\uparrow$\smash{(\%)} & $\downarrow$ & $\downarrow$ & $\uparrow$\smash{(\%)} & $\uparrow$\smash{(\%)} & $\uparrow$\smash{(\%)} \\
\midrule
DeepCAD & 81.12 & 0.921 & 0.995 & 86.24 & 91.7 & 85.8 \\
SkexGen & 83.39 & 0.808 & 0.632 & 86.73 & 99.1 & 99.8 \\
HNC-CAD & 85.33 & 0.772 & 0.558 & 76.07 & 93.9 & 99.7 \\
\midrule
Ours (CE) & 86.67 & 0.764 & 0.616 & 95.88 & 94.5 & 98.6 \\ 
Ours      & 86.67 & 0.768 & 0.593 & 99.65 & 94.4 & 98.3 \\ 
\bottomrule
\end{tabular}
}
\caption{Quantitative comparison of unconditional CAD generation. The second-to-last row reports the results of RECAD with contour extraction. Both MMD and JSD are multiplied by $10^2$.}
\label{tab:model_performance}
\end{table}

\subsubsection{Quantitative evaluation.}
As reported in Tab.~\ref{tab:model_performance}, RECAD outperforms other methods in terms of COV and MMD. 
This suggests that RECAD is capable of producing more diverse parametric CAD models.
While our method achieves a slightly worse JSD score compared to HNC-CAD, it still exceeds SkexGen. RECAD also achieves competitive Novel and Unique scores. Furthermore, our method significantly surpasses all baseline models in terms of Valid, highlighting the robustness of our image-based sketch representation and the effectiveness of our generative framework.

\subsubsection{Qualitative evaluation.}

Fig.~\ref{fig:comparison} presents qualitative comparisons among different methods across varying numbers of extrusion operations.
A key advantage of our image-based representation lies in its ability to capture the complexity of input sketches, which is particularly evident when dealing with a limited number of extrusions. 
With one or two extrusion operations, other methods produce relatively simple models, while RECAD generates more intricate designs.
As the number of extrusions increases, RECAD excels in generating complex sketches while ensuring logical coherence between different extruded parts, maintaining the authenticity of the CAD model.
We include additional results and shape novelty analysis that compares generated models with their closest training samples in the supplementary material.

\begin{figure}[t]
\centering
\includegraphics[width=1\linewidth]{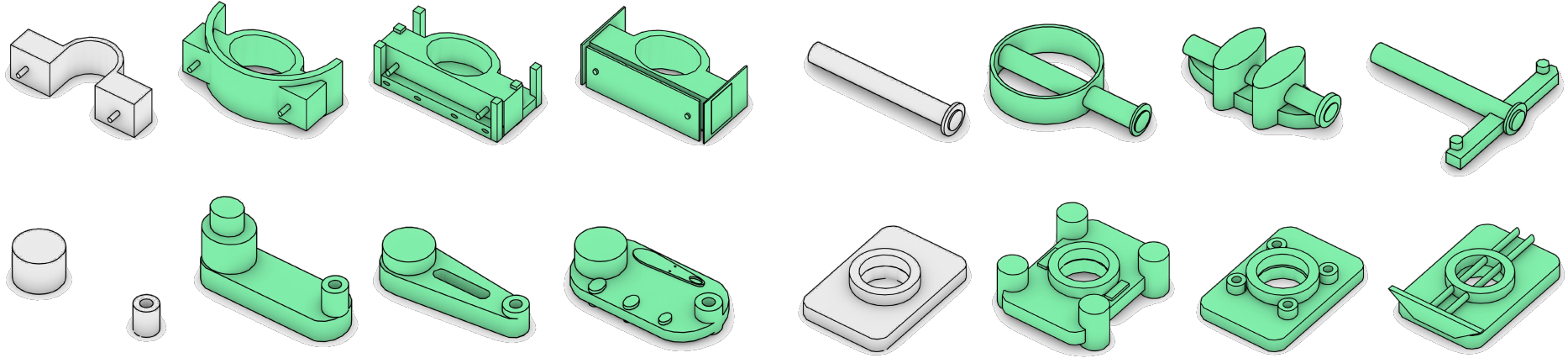}
\caption{Autocompleted CAD models (green) from partial inputs (gray).}\label{fig:autocomplete}
\end{figure}

\subsection{Controllable CAD Generation}

We demonstrate controllable generation results of RECAD in three application scenarios for CAD design. 

\begin{figure}[t]
\centering
\includegraphics[width=1\linewidth]{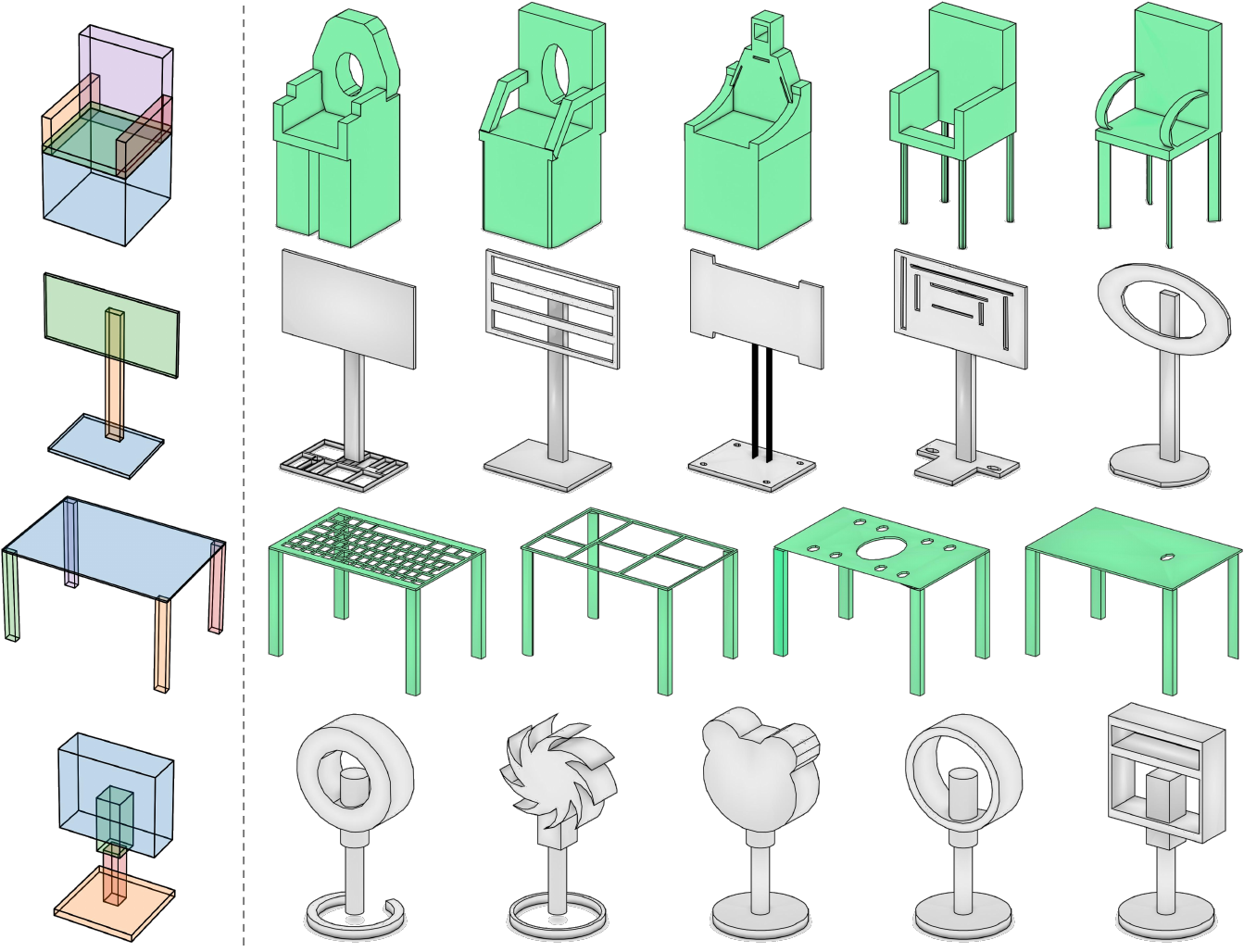}
\caption{Extrusion box conditional generation. Different colors are used for clear visual distinction.}\label{fig:box_condition}
\end{figure}

\subsubsection{Autocompletion from partial input.}
A set of sketch-extrusion pairs with an amount below the extrusion box limit is considered a partial input. Fig.~\ref{fig:autocomplete} shows that our pre-trained model can automatically generate the missing parts based on existing extrusion boxes and sketches, resulting in more complex CAD models. The autocompletion process involves two stages: extrusion box completion and sketch completion. Drawing inspiration from ReShape \cite{lugmayr2022repaint} and BrepGen \cite{xu2024brepgen}, at each time step of the denoising process, we simultaneously diffuse the partial input to the corresponding time step and replace a randomly selected portion of the tokens being denoised with the diffused partial input. The completed extrusion boxes obtained in the first stage are used as conditions to assist in generating sketches in the second stage. 

\subsubsection{Extrusion box conditional generation.}
We observe that the extrusion boxes significantly influence the overall structure of the CAD model, sometimes even suggesting its intended function. Inspired by layout-conditioned image and 3D scene generation \cite{zheng2023layoutdiffusion,bahmani2023cc3d}, we explore CAD generation conditioned on the extrusion boxes. As shown in Fig.~\ref{fig:box_condition}, we manually create extrusion boxes (see the first column) and then use RECAD to generate sketches, resulting in various plausible outcomes.
Interestingly, although trained primarily on mechanical parts from the DeepCAD dataset, our model generated reasonable shapes even when presented with extrusion boxes representing common objects like chairs and tables.
This approach streamlines CAD creation and grants designers efficient, high-level control over the generated model structure.

\begin{figure}[t]
\centering
\includegraphics[width=1\linewidth]{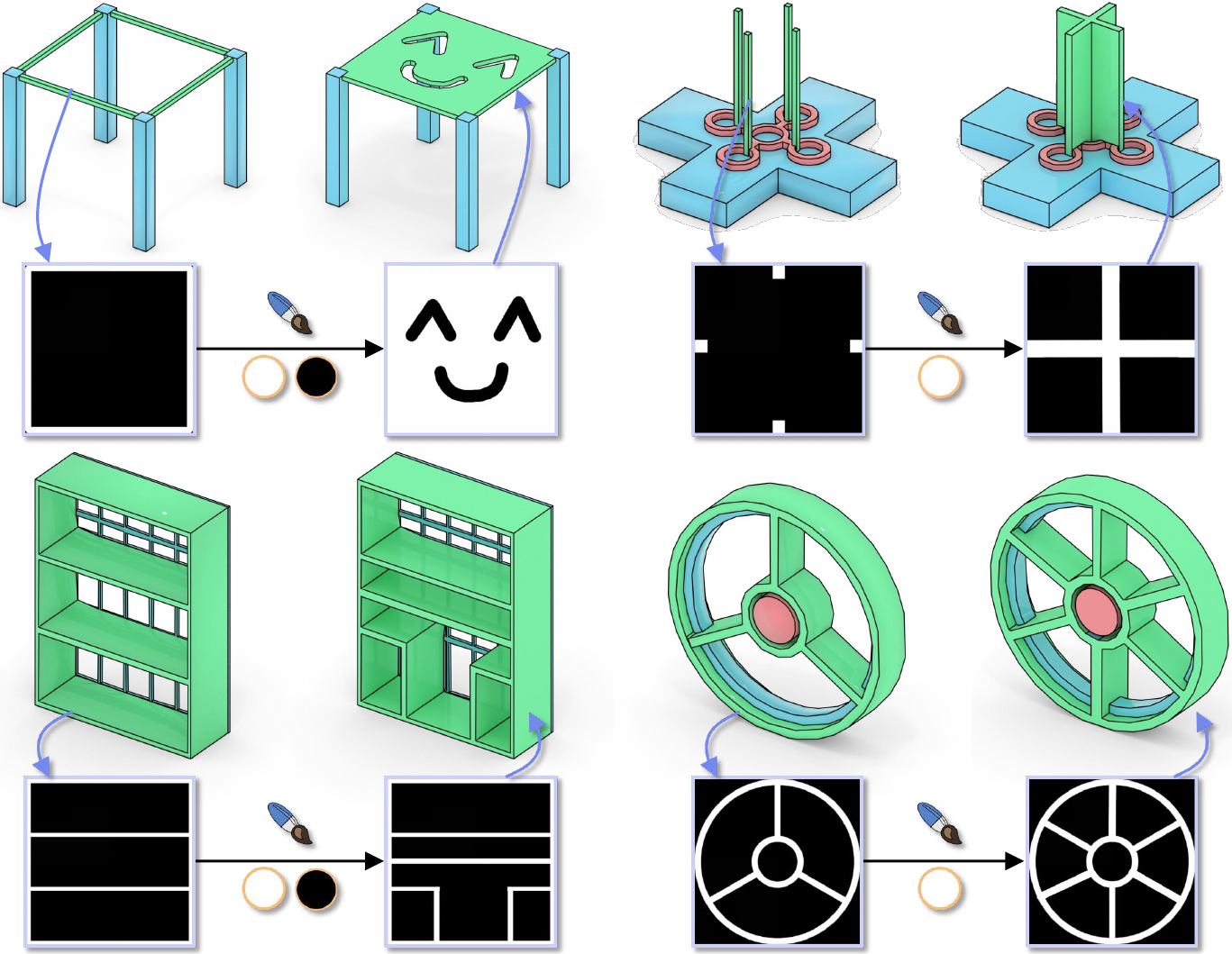}
\caption{Examples of user edits to the generated results. After selecting the target extrusion, users can directly modify the appearance of the 3D model through simple doodling.}\label{fig:edit}
\end{figure}

\subsubsection{Image-based sketch editing.}
A key benefit of employing binary raster sketches in RECAD is the intuitiveness of the editing process. As demonstrated in Fig.~\ref{fig:edit}, users can select the extrusion target and modify the sketch directly with simple black or white brush strokes within any standard image editor. This eliminates the need for specialized CAD software, making 3D CAD model creation and modification remarkably user-friendly.

\section{Limitations and Future Work}

While image-based sketches in RECAD provide a natural interface for design, their dependence on raster images can impede the accurate representation of intricate contours, particularly in low-resolution contexts. Moreover, current CAD model generation methods, including RECAD, lack built-in mechanisms to ensure the plausibility and realism of the generated designs. Introducing adversarial losses or human-in-the-loop feedback mechanisms could be a promising direction for future research. Additionally, given the widespread use of image-based representations over curve sequences, exploring the integration of RECAD with complementary modalities such as text descriptions or semantic segmentation could further enhance the design process.

\section{Conclusion}

We have presented a novel generative diffusion framework for CAD model generation. As the key to our approach, we have introduced a novel sketch-extrusion representation consisting of raster sketch and extrusion box pairs to substitute the predominant sequence-based representation. A two-stage denoising process is adopted to sequentially generate sketch-extrusion pairs which are then converted to CAD models. Extensive quantitative and qualitative evaluations show that RECAD surpasses previous approaches, generating CAD models with greater diversity and complexity. 
Furthermore, RECAD offers users an intuitive and visual interface for modifying both the global topology and local geometry of CAD models, making the editing process more accessible and aligned with user expectations.

\section{Acknowledgments}
This work is partially funded by the National Natural Science Foundation of China (12494550, 12494553, 62172416 and 62172415), the Strategic Priority Research Program of the Chinese Academy of Sciences (XDB0640000 and XDB0640200), the Beijing Natural Science Foundation (Z240002), and the Guangdong Basic and Applied Basic Research Foundation (2023B1515120026).
\bibliography{paper}

\clearpage
\begin{strip}
\begin{center}
\Large\textbf{Supplementary Material}
\end{center}
\hrule
\end{strip}
\vspace{1cm}
\setcounter{page}{1}
\appendix

\input{supp}

\end{document}

%% file: supp.tex
\section{Implementation Details}
The backbone of the extrusion diffusion model features a standard Transformer \cite{vaswani2017attention} architecture with pre-layer normalization, while the sketch denoiser backbone is a 3D conditional UNet with 4 standard layers featuring ResNet blocks and temporal convolutions, 4 cross-attention layers, and a conditioning feature dimension of 1024. To train the diffusion models, input sequences are randomly shuffled to avoid the impact of sequence order. Padding through the duplication of randomly selected sequences is adopted to achieve the predefined amount of extrusions.

\section{t-SNE Embedding of Sketch VAE}

\begin{figure}[!t]
\centering
\vspace{-1cm}
  \includegraphics[width=1\linewidth]{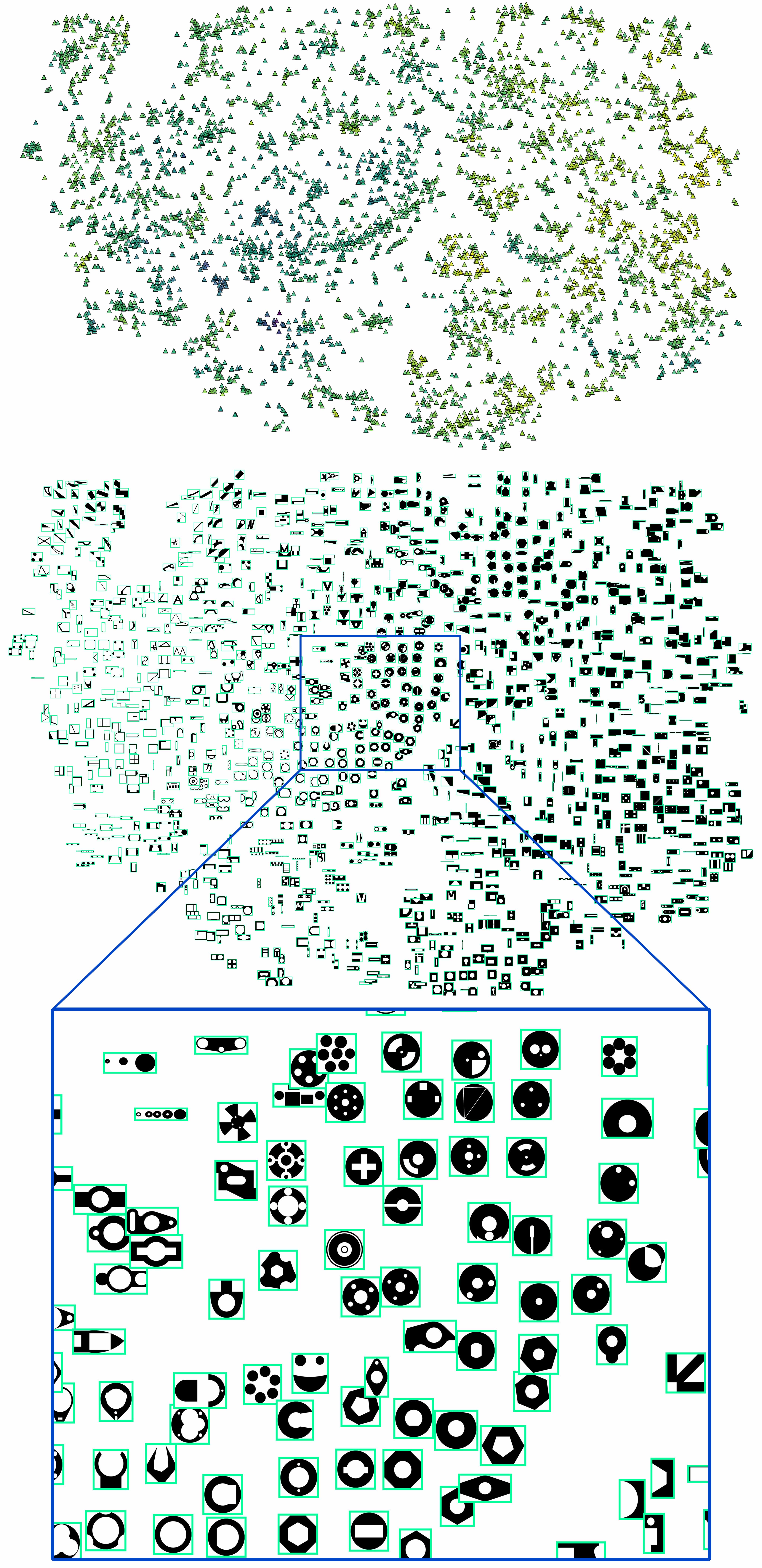}    
  \caption{t-SNE embedding. Please see the text description on the left for details.
  }
  \label{fig:tsne}
\end{figure}

We visualize a 2D t-SNE \cite{van2008visualizing} projection of the latent space learned by the RECAD sketch VAE in Fig. \ref{fig:tsne}. The top portion of the figure illustrates the distribution of the latent feature vectors, with darker triangles indicating larger vector magnitudes. Below, we map these feature vectors to their corresponding input sketches, where black regions represent the profile and white regions represent the background. A blue box highlights a zoomed-in region, offering a more detailed view of the clustering behavior. We observe a clear clustering of geometrically similar sketches in the latent space, demonstrating that our image-based sketch VAE learns meaningful embeddings conducive to downstream tasks in CAD modeling and sketch retrieval.

\section{More Qualitative Results}

\begin{figure*}[t]
\centering
  \includegraphics[width=1\linewidth]{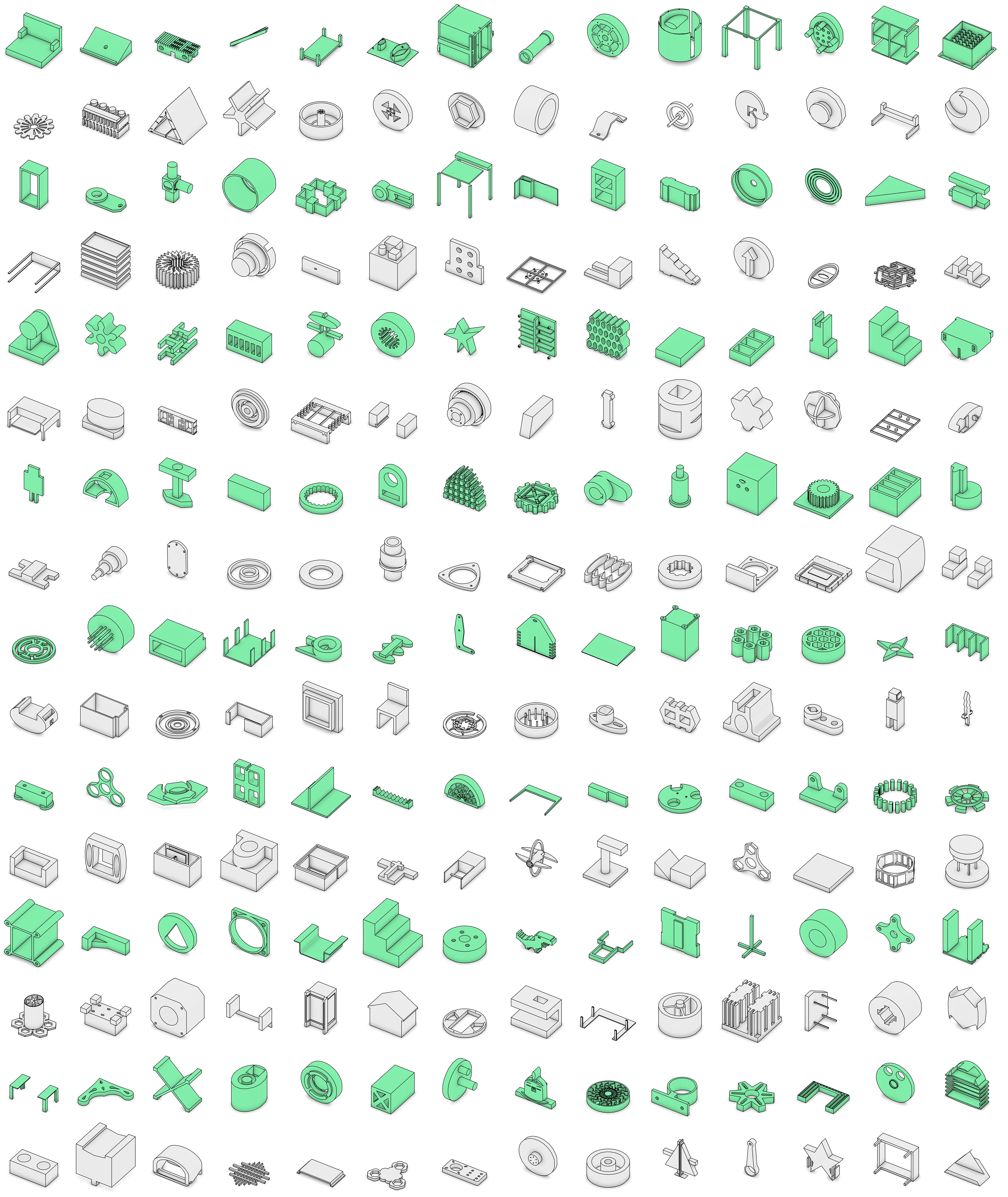}    
  \caption{More examples of CAD models generated unconditionally by RECAD. 
  }
  \label{fig:supp_1}
\end{figure*}

\begin{figure*}[t]
\centering
  \includegraphics[width=1\linewidth]{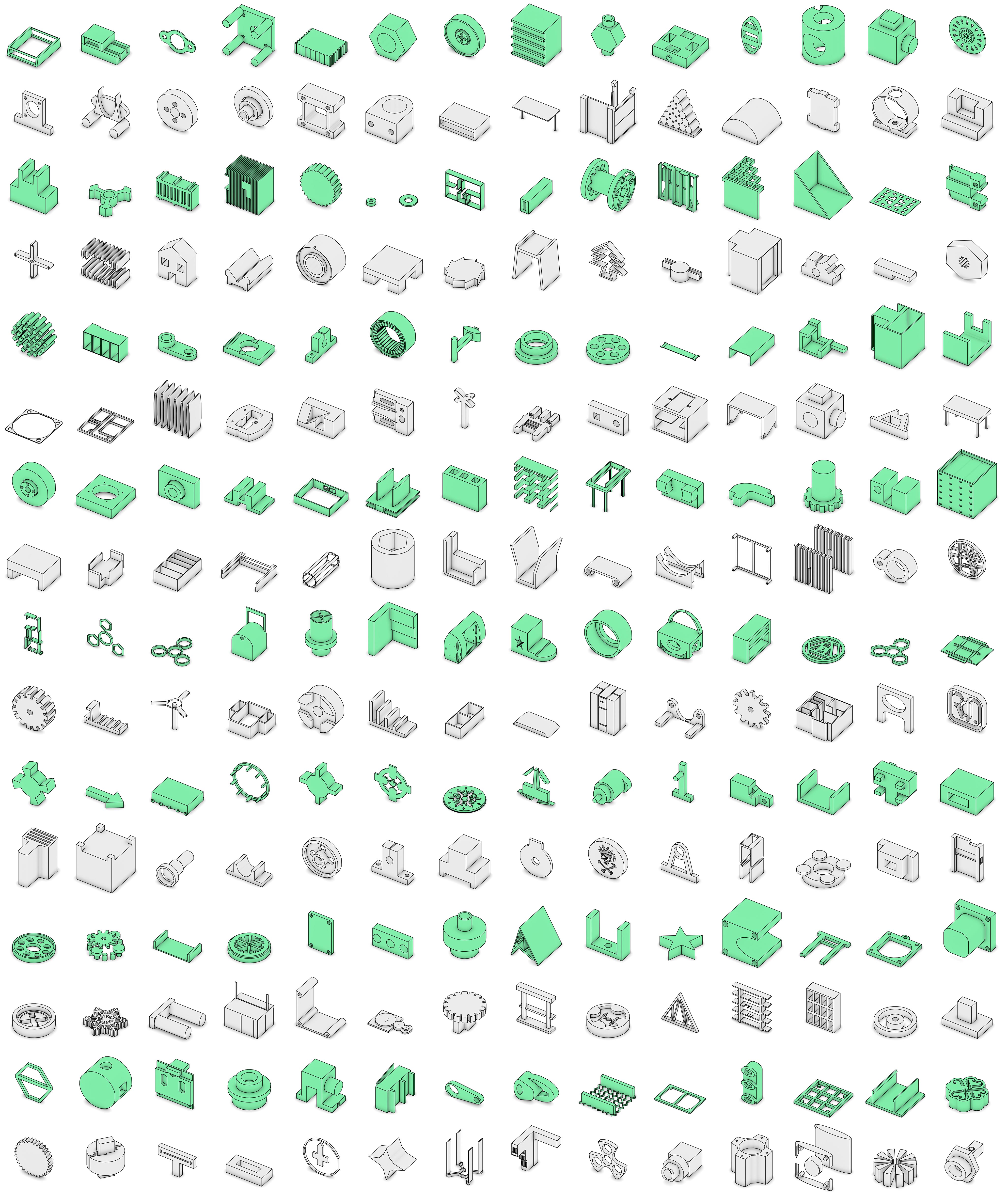}    
  \caption{More examples of CAD models generated unconditionally by RECAD. 
  }
  \label{fig:supp_2}
\end{figure*}

\begin{figure*}[b]
\centering
  \includegraphics[width=1\linewidth]{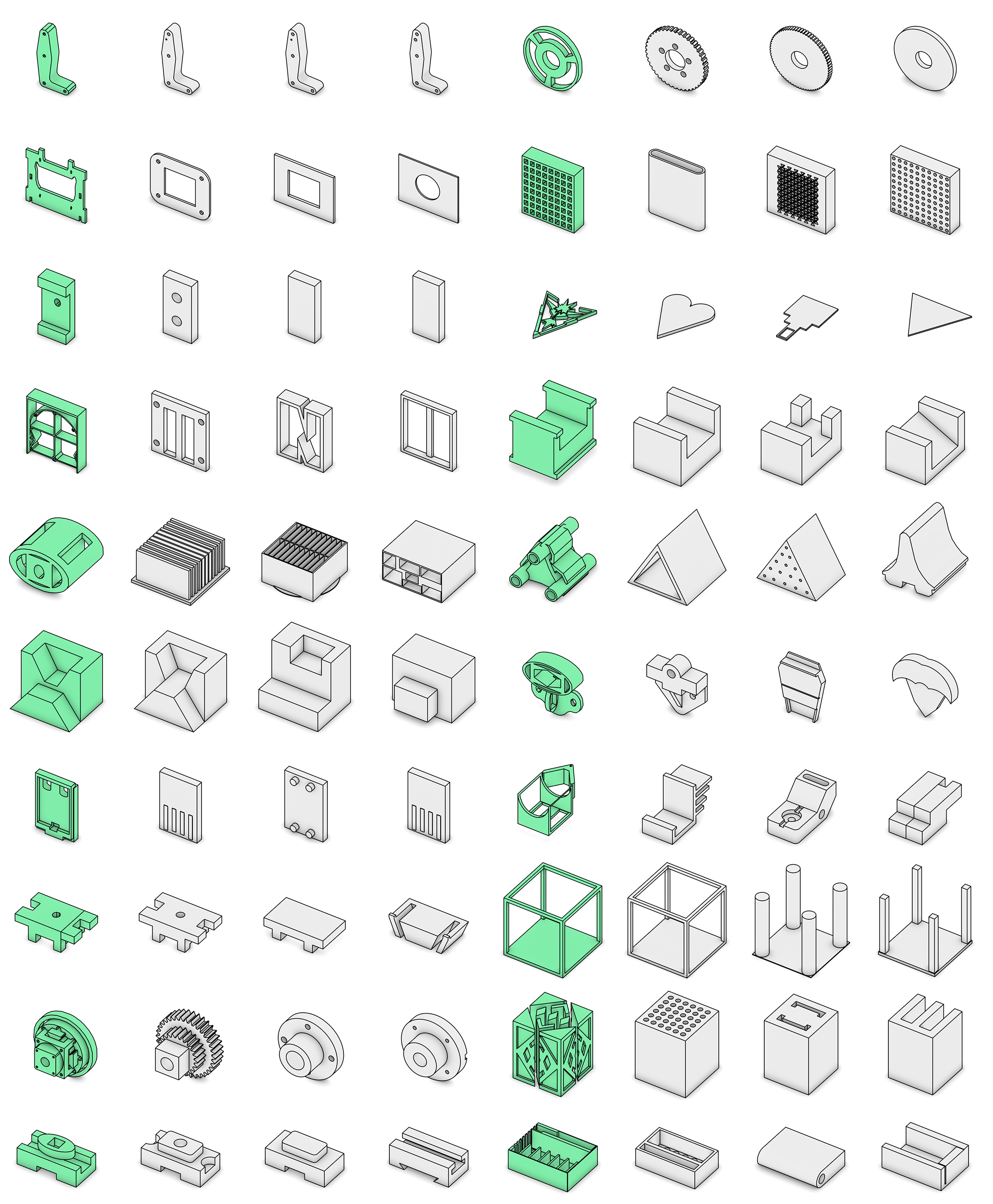}    
  \caption{More examples of CAD models generated unconditionally by RECAD. 
  }
  \label{fig:most_close}
\end{figure*}

This section presents two supplementary analyses for RECAD. We first showcase a comprehensive set of unconditional generation results in Fig. \ref{fig:supp_1} and Fig. \ref{fig:supp_2}, where two distinct colors are used purely for visual clarity. In the second analysis, we examine the novelty of the generated shapes by comparing them with the training set. For all generated examples shown in the main paper, we retrieve their closest counterparts from the training set using Chamfer distance, as illustrated in Fig. \ref{fig:most_close}. The generated shapes (in green) are shown alongside their three nearest neighbors (in white) from the training set. Except for two examples in the third-to-last row, all generated shapes exhibit notable differences from their nearest neighbors, demonstrating that RECAD can produce novel geometric structures beyond mere reproduction of training samples.

